\documentclass[aps,prb,twocolumn,superscriptaddress,floatfix,longbibliography]{revtex4-2}
\usepackage[utf8]{inputenc}
\usepackage{graphicx}
\usepackage{braket}
\usepackage{bm}
\usepackage{acronym}
\usepackage{upgreek}
\usepackage{tabularx}
\usepackage{siunitx}
\usepackage{physics}
\usepackage{booktabs}

\usepackage{hyperref}

\begin{document}
\title{Classical dynamics and semiclassical analysis of excitons in cuprous oxide}
\author{Jan Ertl}
\author{Michael Marquardt}
\author{Moritz Schumacher}
\author{Patric Rommel}
\author{Jörg Main}
\email[Email: ]{main@itp1.uni-stuttgart.de}
\affiliation{Institut für Theoretische Physik I,
  Universität Stuttgart, 70550 Stuttgart, Germany}
\author{Manfred Bayer}
\affiliation{Experimentelle Physik 2, Technische Universität Dortmund,
  44221 Dortmund, Germany}

\date{\today}

\begin{abstract}
Excitons, as bound states of electrons and holes, embody the solid state 
analogue of the hydrogen atom, whose quantum spectrum is explained
within a classical framework by the Bohr--Sommerfeld atomic model.
In a first hydrogenlike approximation the spectra of excitons are also
well described by a Rydberg series, however, due to the surrounding
crystal environment deviations from this series can be observed.
A theoretical treatment of excitons in cuprous oxide needs to include 
the band structure of the crystal, leading to a prominent fine-structure
splitting in the quantum spectra.
This is achieved by introducing additional spin degrees of freedom
into the system, making the existence and meaningfulness of classical
exciton orbits in the physical system a non-trivial question.
Recently, we have uncovered the contributions of periodic exciton
orbits directly in the quantum mechanical recurrence spectra of cuprous
oxide [J.~Ertl \emph{et al.}, Phys.\ Rev.\ Lett.\ \textbf{129}, 067401 (2022)]
by application of a scaling technique and fixing the energy of the
classical dynamics to a value corresponding to a principle quantum
number $n=5$ in the hydrogenlike case. 
Here, we present a comprehensive derivation of the classical and
semiclassical theory of excitons in cuprous oxide.
In particular, we investigate the energy dependence of the exciton dynamics.
Both the semiclassical and quantum mechanical recurrence spectra
exhibit stronger deviations from the hydrogenlike behavior with
decreasing energy, which is related to a growing influence of the
spin-orbit coupling and thus a higher velocity of the secular motion
of the exciton orbits.
The excellent agreement between semiclassical and quantum mechanical
exciton recurrence spectra demonstrates the validity of the classical
and semiclassical approach to excitons in cuprous oxide.
\end{abstract}

\maketitle


\acrodef{PSOS}[PSOS]{Poincar\'e surface of section}
\acrodefplural{PSOS}[PSOS]{Poincar\'e surfaces of section}


\section{Introduction}
\label{sec:introduction}
Since the early days of quantum mechanics and the development of
Bohr's model for the hydrogen atom there has been a long standing
debate on the significance of classical dynamics in quantum theory.
The states of the hydrogen atom can well be described by application
of the semiclassical Bohr--Sommerfeld quantization rules~\cite{Sommerfeld1916},
however, the old quantum theory already fails for the computation of
the ground state energy of the helium atom with its underlying
classically chaotic three-body dynamics.
Modern semiclassical theories for both regular and chaotic
multidimensional systems have been derived in the 1970's;
Gutzwiller's periodic-orbit theory~\cite{Gut90} describes the
density of states of chaotic systems in terms of periodic-orbit
parameters of the underlying classical system, an analogous theory has
been developed by Berry and Tabor for integrable systems~\cite{Berry76}.
The semiclassical trace formulas of these theories are the foundation
for, e.g., the physical interpretation of quantum spectra of the
diamagnetic Kepler problem~\cite{Fri89,Has89,Main1999a}, the helium
atom~\cite{Ezra1991}, and the application of random matrix theory to
the quantum spectra of classically chaotic systems~\cite{Haake2018}.

A system similar to the hydrogen atom occurs in solid state physics.
When an electron is excited from the valence band of a semiconductor
to the conduction band it leaves behind a positively charged hole in
the valence band.
Due to the Coulomb interaction between electron and hole the two
particles can form a bound hydrogenlike state, called exciton. 
While semiclassical approaches are now well established in atomic physics,
the majority of theoretical investigations on excitons are performed
within a quantum mechanical framework~\cite{Schoene2016,SchoeneLuttinger,%
ObservationHighAngularMomentumExcitons,ImpactValence,frankevenexcitonseries}. 
This might be due to historical reasons; when first discovered in 1956
only exciton states with low principle quantum numbers were
experimentally accessible~\cite{Gross1956}.
This situation has changed in 2014 when Kazimierczuk \emph{et al.}\
observed giant Rydberg excitons up to principle quantum numbers $n=25$
in cuprous oxide~\cite{GiantRydbergExcitons}.
Recently even Rydberg excitons with quantum numbers up to $n=30$ could
be resolved~\cite{Versteegh2021}.
Taking into account the material parameters of cuprous oxide the size
of these states is of the order of several $\mu\mathrm{m}$, and thus
the classical correspondence principle should be applicable.

The question whether or not the quantum mechanical exciton spectra can
be explained in terms of a classical exciton dynamics is nontrivial
and interesting due to the fact that the experimental spectra of
cuprous oxide are far more complicated than hydrogenlike Rydberg spectra.
In particular, the exciton spectra exhibit a prominent fine-structure
splitting~\cite{ObservationHighAngularMomentumExcitons}.
In the quantum computations this fine structure can be considered by
introducing additional spin degrees of freedom, viz.\ a quasispin $I$,
which couples to the spin $S_{\mathrm{h}}$ of the holes in the valence
band~\cite{Lipari1977,Uihlein1981,ImpactValence}.
To obtain a classical picture an adiabatic approach
can be applied, where the spin degrees of freedom are treated in a
quantum mechanical framework, while the relative coordinates and
momenta are considered as classical variables~\cite{Ertl2020}.

A semiclassical analysis of quantum spectra is possible by extracting
the periodic orbit parameters of the underlying classical dynamics by
using the semiclassical trace formulas, which describe the density of
states as a superposition of sinusoidal contributions from periodic
orbits~\cite{Berry76,Gut90}.
This is most easy for systems possessing a scaling property in a way
that the classical dynamics does not depend on the energy or a
suitable scaling parameter~\cite{Main1999a}.
Contrary to hydrogenlike systems, which become energy-independent by
application of an appropriate scaling transformation~\cite{Main1999a}, the
dynamics of excitons in cuprous oxide, due to the spin-orbit coupling,
still depends on the energy, however, a scaling property can be
recovered via a modification of the spin-orbit term~\cite{Ertl2022Signatures}.
The Fourier transform of the quantum mechanical density of states then
yields quantum mechanical recurrence spectra with peaks corresponding
to actions of periodic exciton orbits.
In Ref.~\cite{Ertl2022Signatures} we demonstrated the existence of
classical exciton orbits in the recurrence spectra of a system where
the energy for the classical dynamics is fixed to a value corresponding
to a principle quantum number $n=5$ in the hydrogenlike case.

In this paper we derive the classical and semiclassical theory for
excitons in cuprous oxide.
We discuss the energy dependence of classical exciton
orbits and demonstrate the resulting differences in the corresponding
recurrence spectra.
In Sec.~\ref{sec:excitons} we present the Hamiltonian for excitons in
cuprous oxide.
The adiabatic approach leading to a classical description of excitons
and the resulting classical dynamics are discussed in
Sec.~\ref{sec:classical_exciton_dynamics}.
Here we study the energy dependence of the phase space structure via
\acp{PSOS} of orbits in the two distinct symmetry planes.
Furthermore, we discuss stability properties of periodic orbits, as
well as the organization of the action for one- to three-dimensional
periodic orbits.
The relation of classical exciton orbits to the quantum properties of
the crystal can be established by semiclassical methods.
The techniques used in this manuscript are presented in Sec.~\ref{sec:semi}
as well as the calculation scheme of semiclassical amplitudes.
The semiclassical results are then compared to quantum mechanical
calculations.
In Sec.~\ref{sec:quantum} we introduce the techniques used for the
solution of the quantum mechanical problem, and present a detailed
comparison of semiclassical amplitudes and quantum recurrence spectra.
An outlook and conclusion are given in Sec.~\ref{sec:conclusion}.

\section{Excitons in cuprous oxide}
\label{sec:excitons}

Excitons in a semiconductor like cuprous oxide are excitations of the
crystal, where an electron is lifted from one of the valence bands to
the conduction band, leaving behind a hole.
In a simple model, i.e., neglecting the valence band structure, the
electron and hole, due to the Coulomb interaction, can form bound
states similar to the Rydberg series in the hydrogen atom.
The complex valence band structure of cuprous oxide causes
fine-structure splittings in the spectra, which can be considered via
correction terms to the kinetic energy that break the spherical symmetry.
In the case of excitons in cuprous oxide, these terms originate from
the non-parabolic shape of the uppermost valence bands, which belong
to the irreducible representation $\Gamma_5^+$~\cite{Schoene2016,SchoeneLuttinger} 
of the cubic $O_{\mathrm{h}}$ symmetry group~\cite{Koster1963} of the crystal.
The treatment of the three-dimensional space of Bloch functions leads to the
introduction of auxiliary degrees of freedom compared to the hydrogenlike model, i.e.,
the quasispin $\boldsymbol{I}$ in addition to
  the hole spin $\boldsymbol{S}_{\mathrm{h}}$, whose components are
given by the spin matrices for $I=1$ and $S_{\mathrm{h}}=1/2$,
  respectively~\cite{Luttinger55,Luttinger56,Suzuki74}.
The excitons in cuprous oxide can be described using the
Hamiltonian~\cite{ImpactValence}
\begin{equation}
  H = E_{\mathrm{g}} + H_{\mathrm{e}}(\boldsymbol{p}_{\mathrm{e}})
  + H_{\mathrm{h}}(\boldsymbol{p}_{\mathrm{h}},\boldsymbol{\hat{I}},\boldsymbol{\hat{S}}_{\mathrm{h}})
 -\frac{e^2}{4\pi\varepsilon_0\varepsilon|\boldsymbol{r}_{\mathrm{e}}-\boldsymbol{r}_{\mathrm{h}}|}\,.
\label{eq:hamiltonian}
\end{equation}
Here,
\allowdisplaybreaks
\begin{align}
  &H_{\mathrm{e}} (\boldsymbol{p}_{\mathrm{e}})
    = \frac{1}{2m_\mathrm{e}} \boldsymbol{p}_{\mathrm{e}}^2
\label{eq:H_kin_e}
\end{align}
and 
\allowdisplaybreaks
\begin{align}
  &H_{\mathrm{h}} (\boldsymbol{p}_{\mathrm{h}},\boldsymbol{\hat{I}},\boldsymbol{\hat{S}}_{\mathrm{h}})
    = \frac{\gamma_1}{2m_0} \boldsymbol{p}_{\mathrm{h}}^2
   +\frac{1}{2\hbar^2m_0} \big[4\gamma_2\hbar^2\boldsymbol{p}_{\mathrm{h}}^2\nonumber\\[1ex]
  &-6\gamma_2(p^2_{\mathrm{h}1}\boldsymbol{I}^2_1+{\rm c.p.})
    -12\gamma_3(\{p_{\mathrm{h}1},p_{\mathrm{h}2}\}\{\boldsymbol{I}_1,\boldsymbol{I}_2\}+{\rm c.p.})\nonumber\\[1ex]
  &-12\eta_2(p^2_{\mathrm{h}1}\boldsymbol{I}_1\boldsymbol{S}_{\rm h1}+{\rm c.p.})
    +2(\eta_1+2\eta_2)\boldsymbol{p}_{\mathrm{h}}^2(\boldsymbol{I}\cdot\boldsymbol{S}_{\mathrm{h}})\nonumber\\[1ex]
  &-12\eta_3(\{p_{\mathrm{h}1},p_{\mathrm{h}2}\}(\boldsymbol{I}_1\boldsymbol{S}_{\rm h2}
    +\boldsymbol{I}_2\boldsymbol{S}_{\rm h1})+{\rm c.p.})\big] + H_{\mathrm{SO}}
\label{eq:H_kin_h}
\end{align}
are the kinetic energy of the electron and hole at positions
$\boldsymbol{r}_{\mathrm{e}}$ and $\boldsymbol{r}_{\mathrm{h}}$ with
momentum $\boldsymbol{p}_{\mathrm{e}}$ and $\boldsymbol{p}_{\mathrm{h}}$,
respectively.
$\{a,b\}=\frac{1}{2}(ab+ba)$ denotes the symmetrized product, c.p.\
stands for cyclic permutation, and $\gamma_i$ and $\eta_i$ are the
Luttinger parameters.
Eq.~\eqref{eq:H_kin_e} differs from the kinetic energy in the
  vacuum only by a modified electron mass $m_\mathrm{e}$, i.e., the
  conduction band remains a parabola.
  By contrast, the degenerate $\Gamma_5^+$ bands significantly
  deviate from a parabolic shape. 
The terms in Eq.~\eqref{eq:H_kin_h} form the Suzuki-Hensel
Hamiltonian~\cite{Suzuki74}, which includes all corrections up to
quadratic order in $\boldsymbol{p}_{\mathrm{h}}$ that are compatible
with the $O_{\mathrm{h}}$ symmetry of the crystal.
The degrees of freedom of the hole in the $\Gamma_5^+$ bands can
be described by an effective internal angular momentum, viz.\
the quasispin $\boldsymbol{I}$.
The last term in Eq.~\eqref{eq:hamiltonian}
is the screened Coulomb potential with the dielectric
constant $\varepsilon$.
In this work we neglect central-cell corrections, which play a role for
the even exciton states only and have been studied in detail in
Refs.~\cite{frankevenexcitonseries,Rommel2021a,Heckoetter2021a}.

\begin{figure}
  \includegraphics[width=0.9\columnwidth]{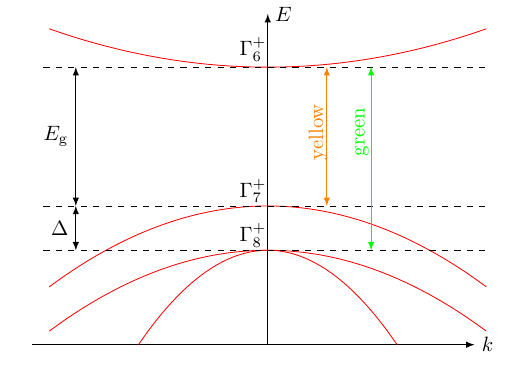}
  \caption{Schematic view of the band structure. The conduction band
    belongs to irreducible representation $\Gamma_6^+$ while at the
    $\Gamma$-point a $\Gamma_7^+$ and $\Gamma_8^+$ valence band
    exists. These bands are separated by the spin-orbit coupling $\Delta$. 
    Transitions from the upper and lower valence band to the
    conduction band result in the yellow and green exciton series,
    respectively.}
  \label{fig:bandstructure}
\end{figure}

When the system is expressed in relative and center-of-mass
coordinates~\cite{Schmelcher1992},
\begin{eqnarray}
  \boldsymbol{r}&= \boldsymbol{r}_{\mathrm{e}}-\boldsymbol{r}_{\mathrm{h}}\, ,\quad
  \boldsymbol{R}=\frac{m_{\mathrm{h}}\boldsymbol{r}_{\mathrm{h}}+m_{\mathrm{e}}\boldsymbol{r}_{\mathrm{e}}}{m_{\mathrm{h}}+m_{\mathrm{e}}}\, ,\nonumber\\
  \boldsymbol{P}&= \boldsymbol{p}_{\mathrm{e}}+\boldsymbol{p}_{\mathrm{h}}\, ,\quad
  \boldsymbol{p}=\frac{m_{\mathrm{h}}\boldsymbol{p}_{\mathrm{e}}-m_{\mathrm{e}}\boldsymbol{p}_{\mathrm{h}}}{m_{\mathrm{h}}+m_{\mathrm{e}}} \, ,
\end{eqnarray}
with vanishing center-of-mass momentum $\boldsymbol{P} = 0$, we obtain
the Hamiltonian~\cite{Lipari1977,Uihlein1981,ImpactValence}
\begin{equation}
  H = E_{\mathrm{g}} + H_{\mathrm{kin}}(\boldsymbol{p},\boldsymbol{I},\boldsymbol{S}_{\mathrm{h}})
  -\frac{e^2}{4\pi\varepsilon_0\varepsilon|\boldsymbol{r}|} + H_{\mathrm{SO}} \, .
\label{eq:H}
\end{equation}
Here the second term,
\begin{align}
  &H_{\mathrm{kin}} (\boldsymbol{p},\boldsymbol{I},\boldsymbol{S}_{\mathrm{h}})
    = \frac{\gamma'_1}{2m_0} \boldsymbol{p}^2
    +\frac{1}{2\hbar^2m_0} \big[4\gamma_2\hbar^2\boldsymbol{p}^2\nonumber\\[1ex]
  &-6\gamma_2(p^2_1\boldsymbol{I}^2_1+{\rm c.p.})
    -12\gamma_3(\{p_1,p_2\}\{\boldsymbol{I}_1,\boldsymbol{I}_2\}+{\rm c.p.})\nonumber\\[1ex]
  &-12\eta_2(p^2_1\boldsymbol{I}_1\boldsymbol{S}_{\rm h1}+{\rm c.p.})
    +2(\eta_1+2\eta_2)\boldsymbol{p}^2(\boldsymbol{I}\cdot\boldsymbol{S}_{\mathrm{h}})\nonumber\\[1ex]
  &-12\eta_3(\{p_1,p_2\}(\boldsymbol{I}_1\boldsymbol{S}_{\rm h2}
    +\boldsymbol{I}_2\boldsymbol{S}_{\rm h1})+{\rm c.p.})\big] \, ,
\label{eq:H_kin}
\end{align}
accounts for the kinetic energy of the electron and hole
quadratic in the momentum $\boldsymbol{p}$,
with $\gamma'_1= \gamma_1 + m_0/m_{\mathrm{e}}$.
Additionally, the Suzuki-Hensel Hamiltonian~\eqref{eq:H_kin_h}
  contains a spherically-symmetric term that couples the
  quasispin and hole spin, viz.\ 
 the spin-orbit term
\begin{equation}
  H_{\mathrm{SO}}=\frac{2}{3}\Delta
  \left(1+\frac{1}{\hbar^2}\boldsymbol{I}\cdot\boldsymbol{S}_{\mathrm{h}}\right)\, .
\label{eq:H_SO}
\end{equation}
Here, $\Delta$ denotes the spin-orbit coupling strength.
This leads to a splitting of the $\Gamma_5^+$ band into a higher lying
two-fold degenerate $\Gamma_7^+$ band, connected to the yellow exciton
series, and a lower lying four-fold degenerate $\Gamma_8^+$ band, connected
to the green exciton series~\cite{Schoene2016,SchoeneLuttinger,Koster1963}.
A schematic of the band structure is shown in Fig.~\ref{fig:bandstructure}.
The first term in the Hamiltonian~\eqref{eq:hamiltonian} is the
band-gap energy $E_{\mathrm{g}}$ between the uppermost $\Gamma_7^+$
valence band and the lowermost $\Gamma_6^+$ conduction band as also
illustrated in Fig.~\ref{fig:bandstructure}.
The material parameters of cuprous oxide are given in Table~\ref{tab:material}.
\begin{table}[b]
    \caption{
        Material parameters of Cu$_2$O used in this paper.
    }
    \label{tab:material}

    \begin{tabular}{llll}
        \toprule
        band-gap energy
        &$E_{\mathrm{g}}$
        &$\phantom{-}\SI{2.17208}{\electronvolt}$
        &\cite{GiantRydbergExcitons}\\
        electron mass
        &$m_{\mathrm{e}}$
        &$\phantom{-}0.99 m_0$
        &\cite{HodbyEffectiveMasses}\\
        hole mass
        &$m_{\mathrm{h}}$
        &$\phantom{-}0.58 m_0$
        &\cite{HodbyEffectiveMasses}\\
        dielectric constant
        &$\varepsilon$
        &$\phantom{-}7.5$
        &\cite{LandoltBornstein1998DielectricConstant}\\
        spin-orbit coupling
        &$\Delta$
        &$\phantom{-}\SI{0.131}{\electronvolt}$
        &\cite{SchoeneLuttinger}\\
        Luttinger parameters
        &$\gamma_1$
        &$\phantom{-}1.76$
        &\cite{SchoeneLuttinger}\\
        &$\gamma_2$
        &$\phantom{-}0.7532$
        &\cite{SchoeneLuttinger}\\
        &$\gamma_3$
        &$-0.3668$
        &\cite{SchoeneLuttinger}\\
        &$\eta_1$
        &$-0.02$
        &\cite{SchoeneLuttinger}\\
        &$\eta_2$
        &$-0.00367$
        &\cite{SchoeneLuttinger}\\
        &$\eta_3$
        &$-0.03367$
        &\cite{SchoeneLuttinger}\\
        \bottomrule
    \end{tabular}
\end{table}

\section{Classical exciton dynamics}
\label{sec:classical_exciton_dynamics}
The classical exciton dynamics of the simple hydrogenlike model with
Hamiltonian
\begin{equation}
  H_{\mathrm{hyd}} = E_{\mathrm{g}} + \frac{\gamma'_1}{2m_0} \boldsymbol{p}^2
  -\frac{e^2}{4\pi\varepsilon_0\varepsilon|\boldsymbol{r}|} \, ,
\label{eq:hamiltonian_hyd}
\end{equation}
is well known.
The bound orbits are classical Kepler ellipses, which are obtained as
analytical solutions of Hamilton's equations of motion.
For the Hamiltonian~\eqref{eq:hamiltonian_hyd} the scaling
\begin{equation}
  \boldsymbol{r} = n_{\mathrm{eff}}^2\tilde{\boldsymbol{r}} \, , \;
  \boldsymbol{p} = \frac{1}{n_{\mathrm{eff}}}\tilde{\boldsymbol{p}} \, ,
\label{eq:r_scal}
\end{equation}
of the coordinates and momenta with the effective quantum number
$n_{\mathrm{eff}}\equiv\sqrt{E_{\mathrm{Ryd}}/(E_{\mathrm{g}}-E)}$,
where $E_{\mathrm{Ryd}}$ is the exciton Rydberg energy, removes the
energy dependence from the Hamiltonian after multiplication with
$n_{\mathrm{eff}}^2$.
This means that, up to a scaling of the Kepler ellipses, the classical
dynamics is the same at all energies.
The situation, however, becomes more complicated when considering the
full Hamiltonian~\eqref{eq:H}.
The additional band structure terms
\begin{align}
  &H_{\mathrm{band}} (\boldsymbol{p},\boldsymbol{I},\boldsymbol{S}_{\mathrm{h}})
    = \frac{1}{2\hbar^2m_0} \big[4\gamma_2\hbar^2\boldsymbol{p}^2\nonumber\\[1ex]
  &-6\gamma_2(p^2_1\boldsymbol{I}^2_1+{\rm c.p.})
    -12\gamma_3(\{p_1,p_2\}\{\boldsymbol{I}_1,\boldsymbol{I}_2\}+{\rm c.p.})\nonumber\\[1ex]
  &-12\eta_2(p^2_1\boldsymbol{I}_1\boldsymbol{S}_{\rm h1}+{\rm c.p.})
    +2(\eta_1+2\eta_2)\boldsymbol{p}^2(\boldsymbol{I}\cdot\boldsymbol{S}_{\mathrm{h}})\nonumber\\[1ex]
  &-12\eta_3(\{p_1,p_2\}(\boldsymbol{I}_1\boldsymbol{S}_{\rm h2}
    +\boldsymbol{I}_2\boldsymbol{S}_{\rm h1})+{\rm c.p.})\big]+H_{\mathrm{SO}}\, ,
\label{eq:H_band}
\end{align}
which are neglected in the hydrogenlike model~\eqref{eq:hamiltonian_hyd},
depend on the additional degrees of freedom introduced by the quasispin
$\boldsymbol{I}$ and the hole spin $\boldsymbol{S}_{\mathrm{h}}$.
Note that the spin-orbit term $H_{\mathrm{SO}}$ given in Eq.~\eqref{eq:H_SO}
disables the scaling procedure described above to remove the energy
dependence of the Hamiltonian~\eqref{eq:H}.

\subsection{Adiabatic approach}
To obtain classical exciton orbits and their parameters from the
Hamiltonian~\eqref{eq:H} the spin degrees of freedom also need to be
considered.
Therefore, we resort to the adiabatic approach introduced in
Ref.~\cite{Ertl2020}.
The idea is based on the different characteristic timescales, related
to the corresponding energy splittings $T\sim \hbar/\Delta E$, on
which the dynamics of the spin degrees of freedom and the relative
coordinates take place.
While for the Rydberg series the splittings $\Delta E\sim 2E_{\mathrm{Ryd}}/n^3$
strongly decrease with increasing quantum number $n$, the spin-orbit
splitting caused by the spin degrees of freedom is fixed to the value
of the spin-orbit coupling $\Delta$.
Comparing the values of $E_{\mathrm{Ryd}}$ and $\Delta$ it becomes
apparent that the dynamics of the spin degrees of freedom becomes much
faster than the dynamics of the relative motion already for quantum
numbers $n \gtrsim 3$. 
This means that the spin degrees of freedom can react almost instantly
to changes in the relative configuration of the coordinates.
Thus we consider the spin degrees of freedom quantum mechanically by
diagonalizing the band-structure part of the Hamiltonian~\eqref{eq:H_band}
in a six-dimensional basis for the quasispin and hole spin $\ket{m_I, m_{S_{\mathrm{h}}}}$
with the corresponding magnetic quantum numbers $m_I$ and $m_{S_{\mathrm{h}}}$.
This yields three distinct two-fold degenerate energy surfaces
$W_k(\boldsymbol{p})$ in momentum space obeying the eigenvalue equation
\begin{equation}
  H_{\mathrm{band}} (\boldsymbol{p},\boldsymbol{I},\boldsymbol{S}_{\mathrm{h}}) \chi_k(\boldsymbol{p}; \boldsymbol{I},\boldsymbol{S}_\text{h} )= W_k(\boldsymbol{p}) \chi_k(\boldsymbol{p}; \boldsymbol{I},\boldsymbol{S}_\text{h} )
\end{equation}
 with the corresponding wave
functions
\begin{equation}
  \chi_k(\boldsymbol{p}; \boldsymbol{I},\boldsymbol{S}_\text{h} )
  =\sum_{m_I,m_{S_\text{h}}}  c_{m_I,m_{S_\text{h}}}(\boldsymbol{p}) \ket{m_I,m_{S_\text{h}}}\, ,
\label{eq:Wavefunction_surface}
\end{equation}
which can be assigned to the different exciton series, i.e.,
one for the yellow series and two for the green series.
The two-fold degeneracies can be explained using Kramers' theorem.
Note that this procedure is somehow the inverse of fitting the
Luttinger parameters $\gamma_i$ and $\eta_i$ to energy surfaces
obtained from spin-DFT calculations~\cite{Schoene2016,SchoeneLuttinger}.
\begin{figure*}
\includegraphics[width=0.98\textwidth]{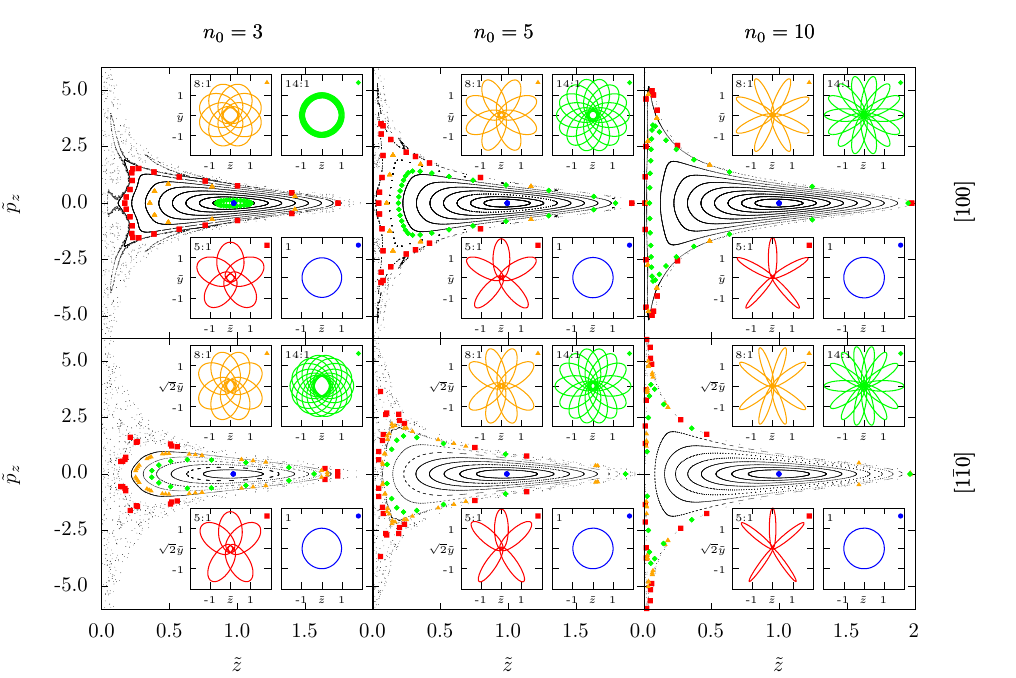}
\caption{\ac{PSOS} for the symmetry planes normal to $[100]$ (top row)
  and normal to $[1\bar{1}0]$ (bottom row) at $n_0=3$ (left), $n_0=5$
  (middle), and $n_0=10$ (right).  A selection of orbits is shown as
  insets, labeled by their winding numbers $M_1=1$ for the
  nearly circular orbits and $M_1{:}M_2$ for the periodic orbits on
  the two-dimensional tori. Their
  positions in the PSOS are marked by corresponding symbols.
  Coordinates and momenta are given in the scaled 
  units~\eqref{eq:r_scal} and thus approximately cover the same range
  for all values of $n_0$.
 The figure extends previous results for $n_0=5$ in the plane
    $\perp [100]$ presented in Ref.~\cite{Ertl2020}.
}
\label{fig:psos_big}
\end{figure*}
The adiabatic approach can now be derived by using the product ansatz
\begin{equation}
  \Psi = \sum_k \Phi_k(\boldsymbol{p})
  \chi_k(\boldsymbol{p}; \boldsymbol{I},\boldsymbol{S}_\text{h})
\label{eq:Wavefunction_full}
\end{equation}
in momentum space.
Inserting this ansatz into the full Hamiltonian~\eqref{eq:H} and
multiplication from the left with $\chi_{k'}$ yields six equations
\begin{equation}
  \boldsymbol{\mathcal{H}} \boldsymbol{\Phi}
  = \left[\left(E_{\mathrm{g}} + \frac{\gamma'_1}{2m_0} \boldsymbol{p}^2
  + W_{k}(\boldsymbol{p})\right)\delta_{k'k}+\boldsymbol{\mathcal{H}}_{\mathrm{C}}
  \right]\boldsymbol{\Phi} = E\boldsymbol{\Phi} \, ,
\end{equation}
where the operator $\boldsymbol{\mathcal{H}}_{\mathrm{C}}$, defined as
\begin{equation}
  (\mathcal{H}_{\mathrm{C}})_{k'k}
  = \bra*{\chi_{k'}}
  \frac{-e^2}{4\pi\varepsilon_0\varepsilon|\boldsymbol{r}|}\ket*{\chi_k} \, ,
\end{equation}
couples the different wave functions $\Phi_k$.
However, for Rydberg excitons with quantum numbers $n \gtrsim 3$ the
coupling terms can be neglected because of the different time scales
of the spin degrees and the relative motion as explained above, which
leads to the Schr\"odinger equation
\begin{equation}
  \left[E_{\mathrm{g}} + \frac{\gamma'_1}{2m_0} \boldsymbol{p}^2
  -\frac{e^2}{4\pi\varepsilon_0\varepsilon|\boldsymbol{r}|}
  + W_{k}(\boldsymbol{p})\right] \Phi_k = E \Phi_k \, .
\end{equation}
In this work we restrict the analysis of the exciton dynamics to the
yellow series described by the classical Hamilton function
\begin{equation}
    \mathcal{H} = E_{\mathrm{g}} + \frac{\gamma'_1}{2m_0} \boldsymbol{p}^2
    -\frac{e^2}{4\pi\varepsilon_0\varepsilon|\boldsymbol{r}|}
    + W_{1,2}(\boldsymbol{p}) =E\, ,
\label{eq:H_class}
\end{equation}
with the two-fold degenerate lowest pair of energy surfaces
$W_{1,2}(\boldsymbol{p})$.
Since the Hamilton function~\eqref{eq:H_class} only depends on the
relative coordinates $\boldsymbol{r}$ and $\boldsymbol{p}$ we obtain
classical exciton orbits by fixing the energy
$E=E_{\mathrm{g}}-E_{\mathrm{Ryd}}/n_{\mathrm{eff}}^2$, i.e., using a
fixed value $n_{\mathrm{eff}}=n_0$, and then numerically integrating
Hamilton's equations of motion
\begin{equation}
  \dot r_i = \frac{\gamma'_1}{m_0} p_i + \frac{\partial
              W_k(\boldsymbol{p})}{\partial p_i} \; , \quad
  \dot p_i = -\frac{e^2}{4\pi\varepsilon_0\varepsilon}
              \frac{r_i}{|\boldsymbol{r}|^3} \, ,
\end{equation}
with, e.g., a standard Runge-Kutta algorithm~\cite{rksuite}.
Due to the additional band structure terms carried by the energy
surfaces $W_k$ the spherical $\mathrm{SO}(4)$ symmetry of the
hydrogenlike problem is reduced to the cubic $\mathrm{O}_\mathrm{h}$
symmetry.
For the cubic symmetry nine symmetry planes exist, where a
two-dimensional motion is possible.
One can distinguish two classes of symmetry planes. 
The three planes normal to the $[100]$ axis and its equivalents
exhibit the same dynamics and likewise the six planes normal to the
$[1\bar{1}0]$ axis and its equivalents.
In contrast to the hydrogenlike model where every starting
configuration leads to a two-dimensional orbit, three-dimensional
orbits are possible when moving the starting configurations out of the
symmetry planes.
This leads to an intricate phase space structure for excitons in
cuprous oxide.

\subsection{Classical exciton orbits and PSOS}
Since the phase space of the classical exciton dynamics described by
the Hamiltonian~\eqref{eq:H_class} is six-dimensional it cannot easily
be visualized.
However, the phase space related to the two-dimensional orbits in the
symmetry planes normal to the $[100]$ and $[1\bar{1}0]$ axes can be
analyzed directly by looking at the corresponding \ac{PSOS}.
They are constructed by choosing a two-dimensional hyperplane in the
four-dimensional phase space, here the $(z,p_z)$ plane, and recording
the intersection points of orbits when crossing the $z$ axis, i.e.,
$x=y=0$.
The remaining momenta $p_x$ and $p_y$ are given by the conservation of
energy and the choice of the symmetry plane.
Such \ac{PSOS} are shown in Fig.~\ref{fig:psos_big} for the two
different symmetry planes and three different values of $n_0$.
In general, the \ac{PSOS} exhibit regular, i.e., integrable or
near-integrable parts of the phase space as toruslike regular
structures, while chaotic motion is indicated by stochastic regions.
Periodic orbits appear as fixed points in the \ac{PSOS}.

\begin{figure}
  \includegraphics[width=\columnwidth]{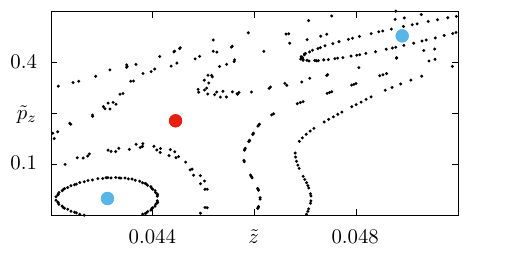}
  \includegraphics[width=\columnwidth]{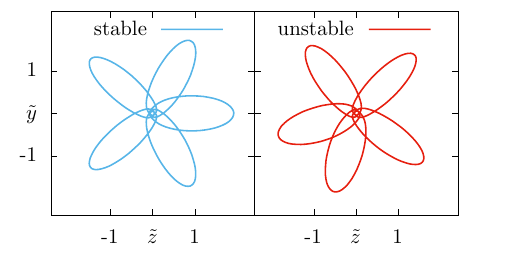}
  \caption{Top: Enlarged part of the \ac{PSOS} for the plane normal to
    $[100]$ at $n_0=5$, where two fixed points are surrounded by
    elliptic and hyperbolic structures, respectively.
    Bottom: Stable and unstable periodic orbit with winding numbers
    $M_1{:}M_2=5{:}1$ corresponding to the two fixed points.}
  \label{fig:psos_SU}
\end{figure}

In all \ac{PSOS} in Fig.~\ref{fig:psos_big} one can observe a central
fixed point, belonging to a nearly circular orbit shown as inset in
the bottom right of the PSOS.
This orbit is surrounded by regular tori, which cover the majority of
phase space.
The outermost parts of the \ac{PSOS} exhibit small stochastic
(chaotic) regions.
The area of the chaotic region is larger for the symmetry plane normal
to $[1\bar{1}0]$ and decreases with increasing values of $n_0$.
In the regular, near-integrable regions the two-dimensional orbits are
characterized as a secular motion of Kepler ellipses.
Here, stable and unstable periodic orbits appear in pairs of elliptic
and hyperbolic fixed points in accordance with the
Poincar\'e--Birkhoff theorem as  illustrated in the enlarged \ac{PSOS}
in Fig.~\ref{fig:psos_SU}.
These periodic orbits can be classified by two integer winding numbers
$M_1$ and $M_2$ with $M_1$ the number of Kepler ellipses and $M_2$ the
number of circulations on the torus caused by the secular motion until
repetition.
Some of them are illustrated in Fig.~\ref{fig:psos_SU} and as insets
in Fig.~\ref{fig:psos_big}.
The assignment of winding numbers can be confirmed by Fourier 
analysis of the periodic orbit coordinate functions~\cite{Gekle2006a,Gekle2007a}.
The ratio $M_1/M_2$ of the winding numbers increases when moving
from the outermost part of the \ac{PSOS} towards the central fixed
point, where it takes its maximum value.
This is related to a decrease of the eccentricity of the Kepler
ellipses and thus to an increase of the angular momentum vector in
the direction perpendicular to the symmetry plane.
The maximum ratio $(M_1/M_2)_{\max}$ is larger in the symmetry
plane normal to $[1\bar{1}0]$ and increases with increasing $n_0$.
It is important to note that the speed of the secular motion strongly
decreases with growing $n_0$ and thus with growing energy.
As can be seen in Fig.~\ref{fig:psos_big} tori with the same ratio
$M_1/M_2$ of the winding numbers are shifted towards the outer
regions of the \ac{PSOS} with increasing $n_0$, which means that
the inner regions more and more belong to tori with higher ratios
$M_1/M_2$ related to orbits with slower secular motion, i.e., the
band structure of the crystal has a stronger influence on states with
low principal quantum numbers and a lower influence on highly excited
Rydberg excitons.
Furthermore, orbits in the two distinct symmetry planes also differ
in their symmetry properties and their orbit parameters.

\begin{figure}[t]
  \includegraphics[width=\columnwidth]{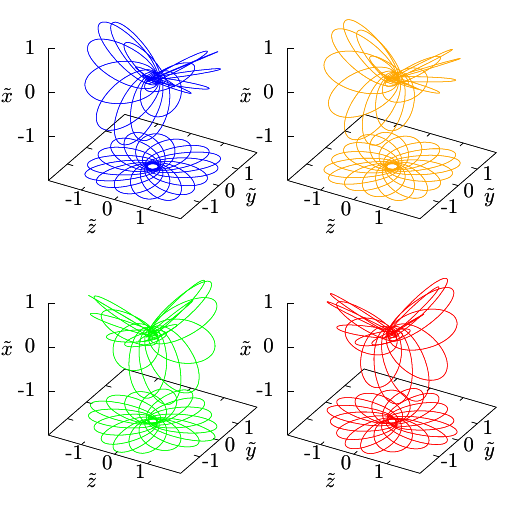}
  \caption{Three-dimensional orbits with winding numbers $M_1{:}M_2{:}M_3=16{:}1{:}2$.
    Four distinct orbits with different orientation and
    position of their maxima exist. The projection onto the $yz$-plane is 
    shown below each orbit.}
  \label{fig:3D_trajectories}
\end{figure}

The exciton dynamics outside the symmetry planes is characterized by
three-dimensional orbits, where a secular motion of Kepler ellipses
occurs in orientations described by two angles $\vartheta$ and $\varphi$.
Here, periodic orbits can be classified by three winding numbers,
where the third winding number $M_3$ counts the cycles of 
the secular motion in the additional direction compared to the 
two-dimensional case.
The orbits appear in sets of four
distinguished variants, not
counting rotations and reflections of the same orbit within the
$O_{\mathrm{h}}$ symmetry group.
A quadruple of three-dimensional orbits with $n_0=5$ and winding
numbers $M_1{:}M_2{:}M_3=16{:}1{:}2$ is illustrated in Fig.~\ref{fig:3D_trajectories}.
The projection of the three-dimensional orbits onto the $yz$ plane
looks similar to the corresponding two-dimensional orbits, where the
orbits in the same column show comparable behavior.
The orbits in the same row have identical orientation towards a
symmetry plane of the crystal.
The orbits in the upper row appear folded towards the plane normal
$[001]$, whereas the orbits in the lower row are oriented towards the
plane normal to $[0\bar{1}1]$.

\subsection{Stability properties of periodic orbits}
The application of semiclassical periodic orbit theories in the
following sections requires the computation of periodic orbit
parameters, including a quantitative description of their stability
properties.
The stability matrix $\boldsymbol{M}(T)$ describes, in a linear
approximation, the deviation of the phase space vector
\begin{equation}
\boldsymbol{\gamma}(t)=\begin{pmatrix} \boldsymbol{r}(t)\\ \boldsymbol{p}(t) \end{pmatrix}
\end{equation}
from the initial point $\boldsymbol{\gamma}(0)$ after one period, i.e.,
\begin{equation}
  \Delta \boldsymbol{\gamma}(T)=\boldsymbol{M}(T) \Delta \boldsymbol{\gamma}(0) \, ,
\end{equation}
and can be calculated by integrating
\begin{equation}
\frac{\mathrm{d}}{\mathrm{d}t} \boldsymbol{M} = \boldsymbol{J}
\frac{\partial^2 H}{\partial \boldsymbol{\gamma} \partial
  \boldsymbol{\gamma}} \boldsymbol{M}, \quad \mathrm{with} \quad
\boldsymbol{J} =\begin{pmatrix} \phantom{-}\boldsymbol{0} & \boldsymbol{1}\\ -\boldsymbol{1} & \boldsymbol{0} \end{pmatrix}
\end{equation}
and $\boldsymbol{M}(0)=\boldsymbol{1}$ along the corresponding orbit.

\begin{figure}
\includegraphics[width=\columnwidth]{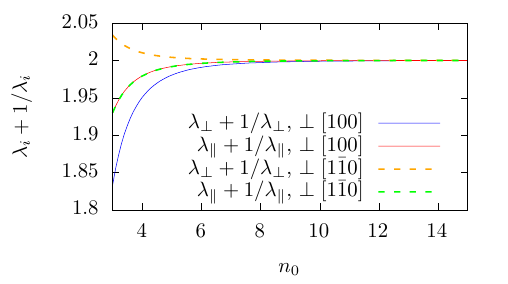}
\caption{Sum $\lambda_i+1/\lambda_i$ for the nearly circular orbits in the
  different symmetry planes, where $\lambda_\parallel$ and $\lambda_\perp$
  describe the stability of the orbits in the plane and out of the
  plane, respectively. The directions normal to the symmetry plane
  are stable for the plane normal to $[100]$ (blue curve) and unstable for the plane 
  normal to $[1\bar{1}0]$ (orange curve). Both orbits are stable with
  respect to perturbations in the plane with almost identical
  stability eigenvalues $\lambda_\parallel$ (red and green curves).}
\label{fig:Stability_circular}
\end{figure}
Since the stability matrix is symplectic the eigenvalues appear in
pairs $\lambda_i$ and $1/\lambda_i$.
An absolute value of the sum $|\lambda_i+1/\lambda_i|>2$ indicates
that the corresponding direction is unstable, while values
$|\lambda_i+1/\lambda_i| \leq 2$ indicate a stable direction.
For every constant of motion a pair of eigenvalues $\lambda_i = 1$
does exist.
The stability eigenvalues also allow us to study the stability
properties of two-dimensional orbits out of the symmetry planes.

\begin{figure}
\includegraphics[width=\columnwidth]{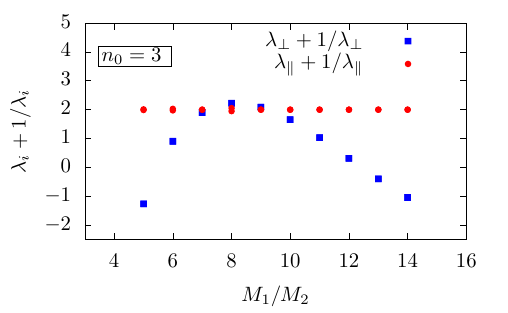}
\includegraphics[width=\columnwidth]{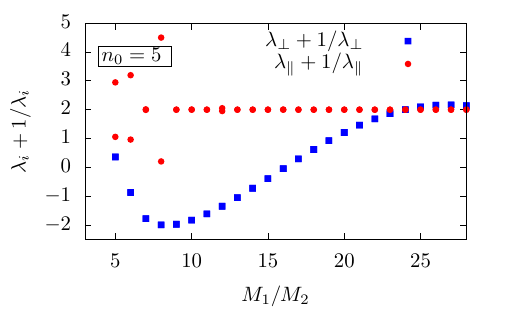}
\includegraphics[width=\columnwidth]{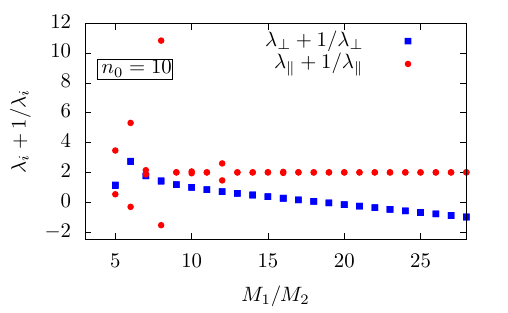}
\caption{Sum $\lambda_i+1/\lambda_i$ for the two-dimensional orbits in the 
  plane normal to $[100]$ for $n_0=3$ (top), $n_0=5$ (middle), and $n_0=10$
  (bottom). $\lambda_\parallel$ describes the stability of the orbits
  in the plane, $\lambda_\perp$ describes the stability of the orbits
  out of the plane.
The stability eigenvalues for $n_0=3$ and $10$ extend results presented for
    $n_0=5$ in the Supplemental Material of Ref.~\cite{Ertl2022Signatures}.}
\label{fig:Stability_plane}
\end{figure}
For the nearly circular orbits the values of $\lambda_i+1/\lambda_i$
as functions of $n_0$ are shown in Fig.~\ref{fig:Stability_circular}.
For both orbits the stability in the symmetry plane (red and green curves)
is almost identical.
For the direction out of the plane the behavior differs.
The nearly circular orbit in the plane normal to $[100]$ is stable and the
orbit normal to $[1\bar{1}0]$ is unstable against perturbations out of 
the symmetry planes.
For the two-dimensional orbits a similar behavior can be observed.
Orbits in the plane normal to $[1\bar{1}0]$ are unstable against 
perturbations of the orbits out of the plane, whereas the orbits in
the symmetry plane normal to $[100]$ are mostly stable.
This can be seen in Fig.~\ref{fig:Stability_plane}, where the sums
$\lambda_i+1/\lambda_i$ for the directions orthogonal to the orbit are
shown for $n_0=3$, $5$, and $10$.
Regarding perturbations within the symmetry plane for a given ratio
$M_1/M_2$ two partner orbits exist, one stable and one unstable, as
already discussed above (see the elliptic and hyperbolic fixed points
in Fig.~\ref{fig:psos_SU}).
The largest deviation from $\lambda_\parallel=1$ is found at low
ratios $M_1/M_2$, where the influence of the band structure terms
on the orbits becomes more pronounced.
The strongest effect occurs for orbits exhibiting high symmetry,
which are the shortest orbits in the fundamental region.

The stability eigenvalues of the two orbits with the same ratio
$M_1/M_2$ against perturbations out of the symmetry plane are nearly
identical and mostly located in the stable region, i.e.,
$|\lambda_\perp+1/\lambda_\perp| \le 2$.
However, for the stability properties in the symmetry plane only the
orbits with a low ratio $M_1/M_2$ show deviations from the
integrable behavior characterized by $\lambda_\parallel=1$ and for
$n_0=3$ no deviations can be observed at all. 
We observe a similar behavior for the periodic orbits in the other
symmetry plane and for the three-dimensional orbits.

\subsection{Action and ordering scheme for orbits in cuprous oxide}
\begin{figure}
  \includegraphics[width=\columnwidth]{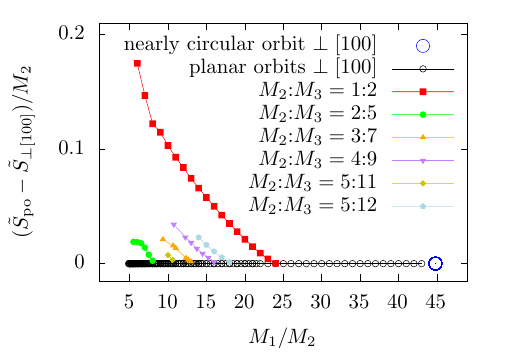}
  \caption{
    Difference of the action $\tilde S_{\mathrm{po}}$ of the periodic orbits (po) at $n_0=5$
    and the corresponding two-dimensional orbits $\tilde S_{\perp[100]}$ 
     in the plane normal to $[100]$ normalized by their winding number
     $M_2$ over
    the ratio of winding numbers $M_1/M_2$. Six different series of
    three-dimensional orbits with winding numbers $M_2{:}M_3$ are shown as 
    solid points, connected by lines to guide the eye.
    The action of two-dimensional orbits in the plane normal to 
    $[100]$ provides a lower border for the families of three-dimensional orbits.
    For increasing ratio $M_1/M_2$ the nearly circular orbit (indicated
    by a blue circle) is reached.}
  \label{fig:S}
\end{figure}
To connect the classical trajectories to quantum properties of the
system the action $S$ of classical orbits is needed.
It can be calculated by integrating the corresponding equation of motion
\begin{equation}
\frac{\mathrm{d}}{\mathrm{d}t} S = \boldsymbol{p} \frac{\mathrm{d} \boldsymbol{r}}{\mathrm{d}t} \,
\end{equation}
along the orbit.
In Fig.~\ref{fig:S} the difference of the action to the one of two-dimenisonal orbits
in the plane normal to $[100]$ normalized by the second winding number $M_2$ is plotted over the ratio $M_1/M_2$ of the first
two winding numbers at $n_0=5$ for selected pairs of winding numbers $M_2{:}M_3$. 
Different values of $n_0$ lead to qualitatively similar pictures.
Note that additional orbits with other pairs of winding
  numbers $M_2{:}M_3$ would fill a finite dense area in Fig.~\ref{fig:S}.
With increasing ratio $M_1/M_2$
the three-dimensional orbits approach the zero line, where the
two-dimensional orbits are located and disappear when this line is reached.
The ratio $M_1/M_2$ where this
is happening for the orbits with $M_2{:}M_3=1{:}2$ corresponds to the region in
Fig.~\ref{fig:Stability_plane} where $|\lambda_i+1/\lambda_i| \to 2$.
A crossing of the line $|\lambda_i+1/\lambda_i|=2$ indicates a change
of stability properties, which is connected to a bifurcation of the
respective orbit. Similary, the eigenvalues of two-dimensional orbits with higher 
values of $M_2$ approach $\lambda_i=1$ where the corresponding 
three-dimensional orbits disappear.
Thus, the two-dimensional orbits constitute a boundary for the
three-dimensional orbits.
For increasing values of the ratio $M_1/M_2$ these orbits approach
the action of the nearly circular orbit in the plane normal to $[100]$
marked by a blue circle at $M_1/M_2 \approx 44.8$ in Fig.~\ref{fig:S}.

\section{Semiclassical approach}
\label{sec:semi}
To reveal the existence of a classical exciton dynamics as described
in Sec.~\ref{sec:classical_exciton_dynamics} the classical dynamics
needs to be connected to the quantum spectra of the crystal.
This connection can be established by application of semiclassical
periodic orbit theory~\cite{Berry76,Tomsovic95,Gut90}.
In these theories, the semiclassical density of states
\begin{equation}
  \varrho_{\mathrm{sc}}(E) = \varrho_0(E)
  + \sum_{\mathrm{po}} {\cal A}_{\mathrm{po}}
  \cos( S_{\mathrm{po}}(E)/\hbar- \frac{\pi}{2}\mu_{\mathrm{po}}) 
\label{eq:rho_sc}
\end{equation}
is given as the sum of the average density of states $\varrho_0(E)$
and a superposition of fluctuations.
The frequencies of the sinusoidal fluctuations depend on the
classical action $S_{\mathrm{po}}$ of the periodic orbits,
the amplitudes ${\cal A}_{\mathrm{po}}$ are determined by their stability
properties, and the Maslov indices $\mu_{\mathrm{po}}$ rely on their
topology. Since in the semiclassical limit the periodic orbit formulas
coincide with the quantum mechanical result, this approach allows for revealing
the contributions of classical orbits to the quantum spectra of the system.
The expression for the semiclassical amplitudes ${\cal A}_{\mathrm{po}}$ 
differs for integrable and non-integrable systems, and the two cases
have to be treated separately.

\subsection{Integrable systems}
\label{sec:integrable_systems}
In an integrable system one can introduce action-angle variables 
\begin{equation}
    J_i=\frac{1}{2\pi} \oint_{{\cal C}_i} \boldsymbol{p} \mathrm{d}\boldsymbol{r}, \quad \vartheta_i=\omega_i t +\beta_i\, ,
\end{equation}
which make the corresponding Hamiltonian independent of the angles $\vartheta_i$.
The action variables $J_i$ therefore provide a set of constants of motion.
When one angle variable goes from $\beta_i$ to $\beta_i+2\pi$ the
system follows an independent irreducible path on a torus ${\cal C}_i$
with frequencies 
\begin{equation} 
\omega_i= \frac{\partial H}{\partial J_i} \, . 
\end{equation}
If all ratios $\omega_i/\omega_j$ for the different frequencies on the 
torus are given by rational numbers the corresponding orbit becomes 
periodic and can be characterized by a set of integer winding numbers
$M_i$, which count the number of cycles on each of the tori ${\cal C}_i$
until the orbit returns to its initial position.
The periodic orbits are located on resonant tori.

For integrable systems the density of states is given by the 
Berry-Tabor formula~\cite{Berry76,Tomsovic95}.
In two dimensions the periodic orbits of such a system can be
characterized by two winding numbers $M_1$ and $M_2$ and
the density of states can be written as~\cite{Tomsovic95}
\begin{align}
\begin{split}
  &\varrho_{\mathrm{sc}}(E) = \varrho_0(E)\\
  &+\frac{1}{\pi \hbar}\sum_{\mathrm{po}}
    \frac{T_{\mathrm{po}}}{\sqrt{\hbar M_2^3 \abs{g''_E}}}
    \cos(S_{\mathrm{po}}/\hbar -\frac{\pi}{2}\mu_{\mathrm{po}} -\frac{\pi}{4}) \, .
\end{split}
\label{eq:BT}
\end{align}
Here, the semiclassical amplitudes depend on the period of the orbit
$T_\mathrm{po}$, the second winding number $M_2$, and on the second
derivative
\begin{equation} 
  g''_E = \frac{\dd^2 g_E}{\dd J_1^2} \, ,
\label{eq:gss}
\end{equation}
of the relation $J_2=g_E(J_1)$ between the two action variables.

\subsection{Non-integrable systems}
When adding a non-integrable perturbation to an integrable system
the resonant tori break up leaving behind isolated periodic orbits.
The contribution of these orbits to the density of states
is captured in Gutzwiller's famous trace formula~\cite{Gut90}
\begin{align}
\begin{split}
  &\varrho_{\mathrm{sc}}(E) = \varrho_0(E)\\
  &+\frac{1}{\pi \hbar}\sum_{\mathrm{po}}
    \frac{T_{\mathrm{ppo}}}{\sqrt{\abs{\det(\boldsymbol{M}_\mathrm{po}-\boldsymbol{1})}}}
    \cos(S_\mathrm{po}/\hbar-\frac{\pi}{2}\mu_{\mathrm{po}}) \, .
\end{split}
    \label{eq:gutz}
\end{align}
In this case, the amplitudes depend on the stability properties of the 
system provided by the monodromy matrix $\boldsymbol{M}_{\mathrm{po}}$,
which describes the linear response of the system to perturbations
in directions orthogonal to the orbit.
The index `ppo' indicates primitive periodic orbits meaning that only
one repetition of the orbit is considered.

\subsection{Systems with scaling property}
The amplitudes in Gutzwiller's trace formula~\eqref{eq:gutz} as well as
in the Berry-Tabor-formula~\eqref{eq:BT} depend on the energy 
or even additional parameters like external fields through the orbit parameters. 
In some systems it is possible to perform a scaling operation in such a way
that the classical orbits no longer depend on a scaling parameter 
\begin{equation}
  w = \frac{1}{\hbar_{\mathrm{eff}}} = \frac{n_{\mathrm{eff}}}{\hbar} \, .
\end{equation}
The action
\begin{equation}
  S_{\mathrm{po}}(w)/\hbar = \tilde S_{\mathrm{po}} w \, ,
\label{eq:scaledAction}
\end{equation}
then only depends linearly on the scaling parameter $w$
with the constant scaled action $\tilde S_{\mathrm{po}}$.
Examples where such scaling techniques have been applied are billiard
systems~\cite{Cvi89} or the hydrogen atom in a magnetic
field~\cite{Hol88,Fri89,Main1999a}.
Transforming the semiclassical density of states~\eqref{eq:rho_sc}
from energy to $w$ domain the resulting expression
\begin{equation}
  \varrho_{\mathrm{sc}}(w) = \varrho_0(w)
  +\Re \sum_{\mathrm{po}} {\cal A}_{\mathrm{po}}
    \exp(i\tilde S_{\mathrm{po}} w) \, ,
\label{eq:rho_series}
\end{equation}
can be understood as a Fourier series with constant periodic orbit 
parameters.
For convenience the Maslov index $\mu_{\mathrm{po}}$ is contained
in a complex valued amplitude ${\cal A}_{\mathrm{po}}$.

The individual periodic orbits provide sinusoidal fluctuations to the 
density of states, which cannot be observed directly.
However, the  contributions of the periodic orbits can be revealed by
Fourier transform from $w$ to the scaled action domain.
The obtained recurrence spectra now exhibit delta peaks at scaled actions 
$\tilde S_{\mathrm{po}}$
\begin{equation}
  C_{\mathrm{sc}}(\tilde S)
  = \sum_{\mathrm{po}} {\cal A}_{\mathrm{po}} \delta(\tilde S - \tilde S_\mathrm{po} )\, ,
\label{eq:rec}
\end{equation}
which allows for a direct assignment of individual orbits to the quantum
mechanical recurrence spectrum.

\subsection{Scaling technique for excitons in cuprous oxide}
Here we apply the scaling technique to the classical exciton orbits in
cuprous oxide.
For all bound states of a hydrogenlike Rydberg spectrum the 
scaling property~\eqref{eq:r_scal} holds. The corresponding
classical orbits are Kepler ellipses, and thus the classical phase
space structure does not depend on the energy of the Rydberg states.
However, the classical dynamics underlying a given exciton state
depends on the energy. 
This can be seen when applying the scaling~\eqref{eq:r_scal} to the 
Hamiltonian~\eqref{eq:hamiltonian}. 
After multiplying by $n_{\mathrm{eff}}^2$
the Hamiltonian reads
\begin{equation}
H=H_{\mathrm{kin}}(\tilde{\boldsymbol{p}},\boldsymbol{\hat{I}},\boldsymbol{\hat{S}}_{\mathrm{h}})
	+n_{\mathrm{eff}}^{2}H_\text{SO}(\boldsymbol{\hat{I}},\boldsymbol{\hat{S}}_{\mathrm{h}})
	-\frac{e^2}{4\pi\varepsilon_0\varepsilon|\tilde{\boldsymbol{r}}|} \, .
\label{eq:H_eff}
\end{equation}
Thus, the impact of the spin-orbit coupling on
the states varies with energy.
This dependence can be avoided by application of a scaling technique
to the spin-orbit coupling.
We apply a scaling technique to the spin-orbit term
$H_{\mathrm{SO}}$ by replacing the coupling constant $\Delta$ in
Eq.~\eqref{eq:H_SO} with an energy-dependent coupling parameter
$\tilde\Delta$, i.e.,
\begin{equation}
\Delta \to \tilde \Delta = \frac{n_0^2}{n_{\mathrm{eff}}^2} \Delta \, ,
\label{eq:Delta_scal}
\end{equation}
where the constant parameter $n_0$ describes the strength of the
scaled spin-orbit coupling.
While changing material parameters as in Eq.~\eqref{eq:Delta_scal}  is not directly possible in an
experiment, it can prove useful in theoretical investigations.
A tunable spin-orbit coupling $\Delta$ has
already been used to study the exchange
interaction in the yellow exciton series~\cite{Rommel2021a}.

\subsection{Calculation of the trace formula amplitudes}
Due to the energy surface $W_1(\boldsymbol{p})$ the dynamics of the 
excitons in cuprous oxide is not integrable. For such systems the density of
states is given by Gutzwiller's trace formula \eqref{eq:gutz}. 
Applying the scaling technique for the spin-orbit coupling~\eqref{eq:Delta_scal}
the amplitudes in Gutzwiller's trace formula read
\begin{equation}
  |{\cal A}_{\mathrm{po}}|=
  \frac{1}{\pi\hbar}
                \frac{\tilde{S}_{\mathrm{ppo}}}{\sqrt{\abs{(\lambda_{\perp}+1/ \lambda_{\perp}-2)(\lambda_{\parallel}+1/ \lambda_{\parallel}-2)}}} \, .
  \label{eq:amp_gutz}
  \end{equation}
Note that in the scaled system the period $T_{\mathrm{ppo}}$ must be replaced by the
scaled action $\tilde{S}_{\mathrm{ppo}}$ \cite{Main1999a}.
For the isolated nearly circular orbits
the stability eigenvalues $\lambda_{\perp}$ and
  $\lambda_{\parallel}$ differ from one, and
Eq.~\eqref{eq:amp_gutz} can be directly
evaluated.
The periodic orbit parameters and amplitudes
for one cycle of these orbits with $n_0=3$, $5$, and $10$
are given in Table~\ref{tab:cent}.
\begin{table}[t]
  \centering
  \caption{
      Periodic orbit parameters and Gutzwiller amplitudes for the
      nearly circular orbits in the planes $\perp$ to $[100]$ and $[1\bar{1}0]$.
  }
  \label{tab:cent}
  
  \begin{tabular}{rccccr
  }
      \toprule
      $n_0$ 
      & plane
      & $\tilde{S}_\mathrm{po}/(2\pi)$ 
      & $\lambda_{\perp} + 1/\lambda_{\perp}$ 
      & $\lambda_{\parallel} + 1/\lambda_{\parallel}$ 
      & $|\mathcal{A}_\mathrm{po}|$
      \\\midrule
                  3 & $[100]$ & 1.0086 &       1.8333 &       1.9295 &       18.6128\\
                  3 & $[1\bar{1}0]$ & 1.0033 &       2.0338 &       1.9299 &       41.2599\\
                  5 & $[100]$ & 0.9983 &       1.9803 &       1.9917 &       157.0315\\
                  5 & $[1\bar{1}0]$ & 0.9965 &       2.0041 &       1.9917 &       343.0293\\
                  10 & $[100]$ & 0.9942 &       1.9988 &       1.9995 &       2619.0137\\  
                  10 & $[1\bar{1}0]$ & 0.9938 &       2.0002 &       1.9995 &       5671.3482\\       
      \bottomrule
  \end{tabular}
\end{table}

\begin{figure}
\includegraphics[width=\columnwidth]{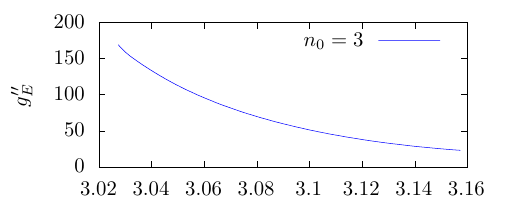}
\includegraphics[width=\columnwidth]{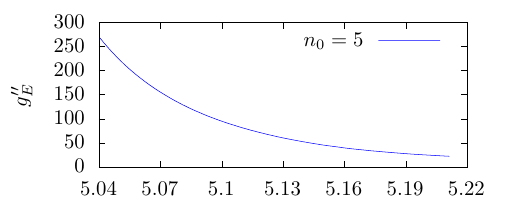}
\includegraphics[width=\columnwidth]{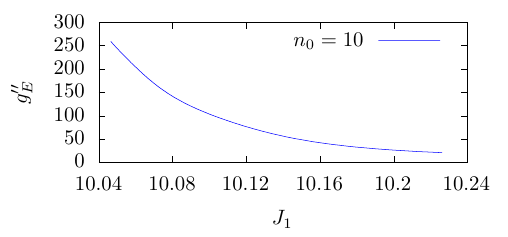}
    \caption{Second derivative of the function $J_2=g_E(J_1)$ with respect to $J_1$
    for the two-dimensional orbits in the plane normal to $[100]$ at $n_0=3$ (top),
    $5$ (middle), and $10$ (bottom).
    The function of $g_E''$ for $n_0=5$ has already been presented in the
      Supplemental Material of Ref.~\cite{Ertl2022Signatures}.}
    \label{fig:gss}
\end{figure}

For the calculation of the amplitudes of the two- and
three-dimensional orbits Eq.~\eqref{eq:amp_gutz}, however, is not
applicable since the majority of orbits exhibit eigenvalue pairs
close to $\lambda=1$, which would lead to the divergence of the amplitude.
For many orbits we find one eigenvalue pair with
$\lambda$ (and thus $1/\lambda$) close to one.
The corresponding degree of freedom can be handled by application of
the Berry-Tabor formula~\eqref{eq:BT}.
The second eigenvalue pair $\lambda_{\mathrm{po}}$ significantly differs from
$\lambda_{\mathrm{po}}=1$, and here the corresponding degree of freedom can be
handled by application of Gutzwiller's trace formula~\eqref{eq:gutz}.
When combining the two semiclassical expressions we arrive at the
semiclassical amplitude
\begin{equation}
|{\cal A}_{\mathrm{po}}|=
\frac{1}{\pi\hbar}
      \frac{1}{\sqrt{\abs{\lambda_{\mathrm{po}}+1/ \lambda_{\mathrm{po}}-2}}}
      \frac{\tilde{S}_{\mathrm{po}}}{\sqrt{\hbar M_2^3 \abs{g''_E}}} \, .
\label{eq:amp_sc}
\end{equation}
The calculation of the semiclassical amplitudes~\eqref{eq:amp_sc}
requires the knowledge of the function $g''_E$ discussed in
Sec.~\ref{sec:integrable_systems}.
For the two-dimensional orbits the action variables $J_1$ and $J_2$
defining the function $J_2=g_E(J_1)$ are constructed with the help of
derivatives of the classical action
\begin{equation}
S_{\boldsymbol{M}}=2\pi (M_1 J_1 + M_2 J_2) \, ,
\label{eq:Action_int}
\end{equation}
with respect to the respective winding number $M_i$.
The derivatives are obtained numerically via difference quotients of
periodic orbits with consecutive winding numbers.
For the three-dimensional orbits an effective two-dimensional
description can be obtained by combining the contributions of the
secular motion in $\varphi$- and $\vartheta$-direction  
described by the greatest common divisor $\tilde{M}_2=\mathrm{GCD}(M_2,M_3)$
giving the action variable 
$\tilde{J}_2=(M_2/\tilde M_2) J_2 + (M_3/\tilde M_2) J_3$.
With the two action variables at hand the function $g''_E$ is obtained
with Eq.~\eqref{eq:gss} by differentiating $J_2$ two times with respect to $J_1$.
In Fig.~\ref{fig:gss} this is illustrated for the two-dimensional orbits in the plane
normal to $[100]$ at $n_0=3$, $5$, and $10$.
The periodic orbit parameters and semiclassical amplitudes of some of
these periodic orbits with $n_0=5$ and winding number $M_2=1$ are given in
Table~\ref{tab:M2_1}.
\begin{table}
  \centering
  \caption{
      Periodic orbit parameters and amplitudes of selected
      two-dimensional orbits in the plane normal to $[100]$ with $M_2=1$ at $n_0=5$.
      The flag s/u denotes the stable or unstable partner orbit.
      For $M_1>12$ the parameters of the stable and unstable orbit are
      numerically identical.}
  \begin{tabular}{
          rc
          S[table-format=3.6, round-precision=6, round-mode=places]
          S[table-format=1.3, round-precision=3, round-mode=places]
          S[table-format=3.3, round-precision=3, round-mode=places]
          S[table-format=3.3, round-precision=3, round-mode=places]}
      \label{tab:M2_1}\\\toprule
      $M_1$
      &{s/u}
      &{$\tilde S_\mathrm{po}/(2\pi)$}
      &{$\lambda_{\perp}+\lambda_{\perp}^{-1}$}
      &{$g''_E$}
      &{$\abs{\mathcal{A}_\mathrm{po}}$}
      \\\midrule
         5 &{s} &      5.59264014 &      0.36122255 &     22.62664747 &      1.65879320\\
         5 &{u} &      5.59266402 &      0.36423413 &     22.62664747 &      1.65879320\\
         6 &{s} &      6.63092273 &      -0.86221512 &     33.28319073 &      1.24074262\\
         6 &{u} &      6.63100811 &      -0.87696864 &     33.28319073 &      1.24074262\\
         7 &{s} &      7.66317907 &      -1.77049867 &     46.19098401 &      1.07192441\\
         7 &{u} &      7.66317908 &      -1.77332297 &     46.19098401 &      1.07192441\\
         8 &{s} &      8.69082391 &      -1.98782913 &     60.43526196 &      1.04156316\\
         8 &{u} &      8.69117999 &      -1.98944048 &     60.43526196 &      1.04156316\\
         9 &{s} &      9.71549473 &      -1.97111319 &     75.35428548 &      1.05175631\\
         9 &{u} &      9.71549479 &      -1.97017316 &     75.35428548 &      1.05175631\\
         10 &{s} &     10.73730967 &     -1.82715032 &     90.87469482 &      1.08390016\\
         10 &{u} &     10.73730969 &     -1.82715305 &     90.87469482 &      1.08390016\\
         11 &{s} &     11.75690885 &     -1.61199237 &    106.56237793 &      1.13331775\\
         11 &{u} &     11.75690881 &     -1.61199497 &    106.56237793 &      1.13331775\\
         12 &{s} &     12.77461808 &     -1.34666196 &    122.21380615 &      1.19927937\\
         12 &{u} &     12.77462250 &     -1.34700787 &    122.21380615 &      1.19927937\\
         13 &{s,u} &     13.79069393 &     -1.04669234 &    138.63540649 &      1.27846537\\
         14 &{s,u} &     14.80532563 &     -0.72308356 &    155.79464722 &      1.37369226\\
         15 &{s,u} &     15.81867394 &     -0.38591419 &    173.52291870 &      1.48982333\\
         16 &{s,u} &     16.83086837 &     -0.04421106 &    191.68707275 &      1.63346211\\
      \bottomrule
  \end{tabular}
\end{table}

\section{Verification of exciton orbits in quantum spectra}
\label{sec:quantum}
Exciton spectra described by the Hamiltonian~\eqref{eq:H} have already
been investigated experimentally \cite{GiantRydbergExcitons}
and theoretically \cite{ImpactValence,frankevenexcitonseries,Rommel2020Green}. 
Here, we want to reveal the existence of classical exciton orbits in
quantum spectra of the yellow exciton series of cuprous oxide.
For this aim we now exploit the scaling property introduced in
Sec.~\ref{sec:semi} by using the scaled spin-orbit
splitting~\eqref{eq:Delta_scal} in quantum computations.
The semiclassical analysis of Fourier transform quantum recurrence spectra then
allows for the observation of signatures of classical exciton orbits and
a detailed study of the energy dependence of the exciton dynamics in the quantum spectra.

\subsection{Scaled exciton spectra}
\label{sec:scaled_quantum}
For the quantum mechanical description of the scaled system
obtained by replacing the spin-orbit coupling in Eq.~\eqref{eq:H_eff} with the scaled version~\eqref{eq:Delta_scal}
we need to 
find the expression for the operators in the scaled coordinates~\eqref{eq:r_scal}.
In quantum mechanics the components of coordinates and momenta 
must satisfy the canonical commutation relations
\begin{equation}
  [ \hat{r}_i, \hat{p}_j ] = i \hbar \delta_{ij} \, .
\label{eq:com}
\end{equation}
Inserting the scaled variables~\eqref{eq:r_scal} into
Eq.~\eqref{eq:com} yields the commutation relations
\begin{equation}
  [ \hat{\tilde r}_i, \hat{\tilde p}_j ] = i \frac{\hbar}{n_{\mathrm{eff}}} \delta_{ij} \, ,
\label{eq:com_scal}
\end{equation}
in the scaled coordinates, where now the Planck constant is replaced
by an effective Planck constant
\begin{equation}
\hbar_{\mathrm{eff}} = \hbar /n_{\mathrm{eff}} \, .
\label{eq:Planck_scal}
\end{equation}
The operators in coordinate space then take the form
\begin{equation}
\hat{\tilde{\boldsymbol{r}}} = \tilde{\boldsymbol{r}}\, , \;
\hat{\tilde{\boldsymbol{p}}} = -i\hbar_{\mathrm{eff}}\nabla_{{\tilde r}}\, .
\label{eq:r_scal_operator}
\end{equation}
In the scaled picture the Schrödinger equation can now be transformed
to the generalized eigenvalue problem
\begin{align}
  &\left[\frac{e^2}{4\pi\varepsilon_0\varepsilon|\tilde{\boldsymbol{r}}|} -n_0^{2}H_\text{SO}(\boldsymbol{\hat{I}},\boldsymbol{\hat{S}}_{\mathrm{h}}) - E_\mathrm{Ryd} \right] |\Psi \rangle \nonumber \\ 
  &= \frac{\hbar^2}{n_{\mathrm{eff}}^2} H_{\mathrm{kin}}
  (-i \nabla_{\tilde r},\boldsymbol{\hat{I}},\boldsymbol{\hat{S}}_{\mathrm{h}}) 
  |\Psi \rangle
\label{eq:H_scal}
\end{align}
for the effective Planck constant
$\hbar_{\mathrm{eff}}=\hbar/n_{\mathrm{eff}}$ (or the effective quantum 
number $n_{\mathrm{eff}}$).
The classical dynamics does not depend on the Planck constant,
which means that for a given $n_0$ the classical dynamics is the same
for all eigenvalues $n_{\mathrm{eff},i}$.
This allows us to reveal contributions of the classical exciton dynamics
to the quantum mechanical recurrence spectra obtained via Fourier transform
of the scaled exciton spectra.

\subsubsection{Matrix representation of the scaled generalized eigenvalue problem}
To obtain the quantum mechanical scaled exciton spectra we need to
solve the scaled Schrödinger equation~\eqref{eq:H_scal}.
To this end we use a complete set of basis functions, which, in
addition to the coordinate wave function, also needs to incorporate the
quasispin $\boldsymbol{I}$ and hole spin $\boldsymbol{S}_{\mathrm{h}}$
degrees of freedom.

Our ansatz for the angular part of the basis states is as follows.
We first couple the quasispin $\boldsymbol{I}$ and the hole spin
$\boldsymbol{S}_{\mathrm{h}}$ to the effective hole spin $\boldsymbol{J}$.
This is an approximate quantum number near the $\Gamma$ point of the
crystal, differentiating between the yellow $J=1/2$ and green $J=3/2$
states.
In a second step, $\boldsymbol{J}$ is combined with the orbital
angular momentum $\boldsymbol{L}$ to form the total angular momentum
$\boldsymbol{F}$.
Without central-cell corrections the electron spin is a good quantum
number of the Hamiltonian and need not be included into our basis.
This makes our coupling scheme slightly different compared to previous
work~\cite{ImpactValence,frankevenexcitonseries}.
As a complete and discrete set of the radial basis functions, we use
the Coulomb-Sturmian
functions~\cite{SturmCommutationRecursion,CoulombSturmForNuclear}
\begin{equation}
  U_{NL}(\rho) = N_{NL}(2\rho)^L\mathrm{e}^{-\rho}L_N^{2L+1}(2\rho) \, ,
\label{eq:basis_r}
\end{equation}
with the radial quantum number $N$ and the dilated radius
$\rho = r/\alpha$.
The parameter $\alpha$ can be used to optimize the convergence
properties of the basis.
With the projection $M_F$ of the total angular momentum
$\boldsymbol{F}$ onto the $z$ axis, we obtain the basis states $|\Pi
\rangle = | N, L, J, F, M_F \rangle$.
When expanding the exciton wave function $|\Psi \rangle$ as
\begin{equation}
  |\Psi \rangle = \sum_{\Pi} c_{\Pi} |\Pi \rangle\, ,
\label{eq:BasisExpression}
\end{equation}
the scaled Schrödinger equation~\eqref{eq:H_scal} can be expressed as
a generalized matrix eigenvalue problem,
\begin{equation}
  \boldsymbol{A}\boldsymbol{c} = \lambda \boldsymbol{B} \boldsymbol{c} \, ,
  \label{eq:generalized_eigenvalue_problem}
\end{equation}
with the matrices
\begin{align}
  \label{eq:MatrixA}
  \boldsymbol{A}_{\Pi'\Pi} &=
  \langle \Pi' |\frac{e^2}{4\pi\varepsilon_0\varepsilon|\tilde{\boldsymbol{r}}|} -n_0^{2}H_\text{SO}(\boldsymbol{\hat{I}},\boldsymbol{\hat{S}}_{\mathrm{h}}) - E_\mathrm{Ryd}| \Pi \rangle \, ,\\
  \label{eq:MatrixB}
  \boldsymbol{B}_{\Pi'\Pi} &= \langle \Pi' | H_{\mathrm{kin}}
  (-i \nabla_{\tilde r},\boldsymbol{\hat{I}},\boldsymbol{\hat{S}}_{\mathrm{h}}) 
  | \Pi \rangle \, ,
\end{align}
and the vector $\boldsymbol{c}$ containing the coefficients $c_\Pi$.
The generalized eigenvalue problem~\eqref{eq:generalized_eigenvalue_problem}
can be solved numerically by using a LAPACK routine~\cite{arpackuserguide}
and a finite number of basis states to obtain a limited number of 
converged eigenvalues $\lambda_i = \hbar^2/n^2_{\mathrm{eff},i}$, and
thus a spectrum with discrete values $n_{\mathrm{eff},i}$ of the
effective quantum number.
The Hamiltonian~\eqref{eq:H} is symmetric under operations of the
cubic group $O_{\mathrm{h}}$.
In particular, this includes a fourfold rotational symmetry around the
$z$ axis, which coincides with our chosen quantization axis.
Because of this, the matrices~\eqref{eq:MatrixA} and
\eqref{eq:MatrixB} have a block diagonal form.
There are four blocks, which are characterized by the magnetic quantum
number $M_\mathrm{F}$ modulo 4 taking the values $1/2$, $3/2$, $5/2$,
and $7/2$, respectively.
Each of these blocks additionally splits into two blocks characterized
by their parity.
We exploit this block structure to accelerate the numerical
calculations.

\subsubsection{Quantum recurrence spectra}
\begin{figure*}
  \includegraphics[width=0.85\textwidth]{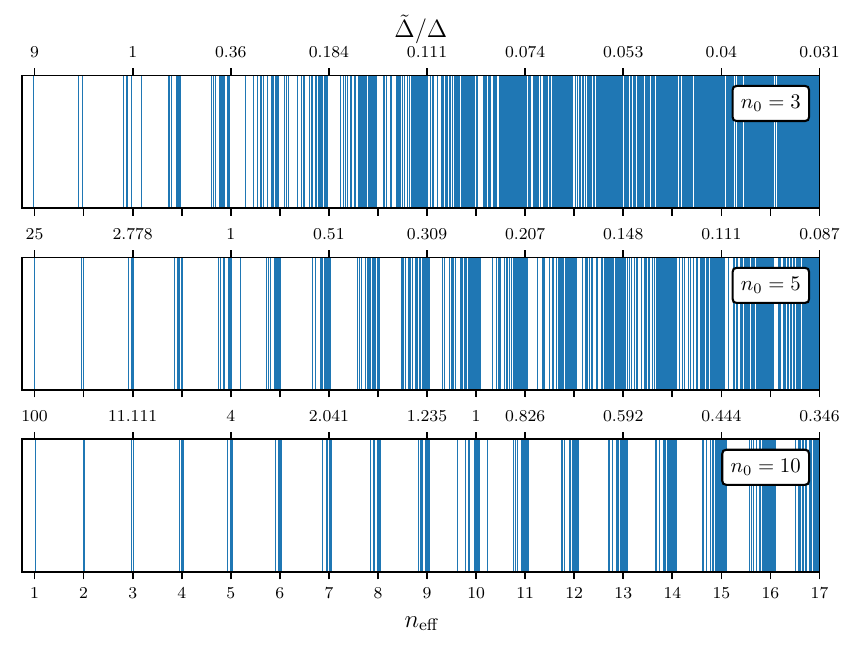}
  \caption{
    Eigenvalue spectra for $n_0=3$ (top), $n_0=5$ (middle), and $n_0=10$ (bottom).
    Each eigenvalue contributes a delta peak to the quantum mechanical 
    density of states~\eqref{eq:scaled_quantum_spectrum}.
    The lowest axis label for the effective quantum number
    $n_{\mathrm{eff}}$ is valid for all three parts of the figure.
    The three upper labels for the ratio $\tilde\Delta/\Delta$ of the
    scaled and real physical spin-orbit coupling are valid for the
    corresponding individual plots.
    Part of the spectrum for $n_0=5$ has already been presented in Ref.~\cite{Ertl2022Signatures}.}
  \label{fig:spectra}
\end{figure*}

We now want to uncover the contributions of classical orbits directly
in the scaled quantum spectra
\begin{equation}
  \varrho(n_{\mathrm{eff}}) = \sum_i  \delta(n_{\mathrm{eff}}-n_{\mathrm{eff},i}) \, .
\label{eq:scaled_quantum_spectrum}
\end{equation}
The eigenvalues of the scaled Schr\"odinger
  equation~\eqref{eq:H_scal} are shown in
Fig.~\ref{fig:spectra} for $n_0=3$, $5$, and $10$, where $n_0$
parameterizes the scaled classical dynamics, as shown in the
\acp{PSOS} in Fig.~\ref{fig:psos_big}, i.e., 
spectra with increasing $n_0$ are related to a classical exciton
dynamics with slower secular motion of orbits.
 The lowest axis label in Fig.~\ref{fig:spectra} gives the eigenvalues
$n_{\mathrm{eff}}$ and the upper labels of the individual plots
compare the scaled spin-orbit coupling $\tilde\Delta$ to the real
physical value $\Delta$, i.e., $\tilde\Delta/\Delta=1$ belongs to the
crystal with the real (unscaled) material parameters of cuprous oxide.

As outlined in Sec.~\ref{sec:semi} the scaled density of
states~\eqref{eq:scaled_quantum_spectrum} can be approximated by a
superposition of sinusoidal fluctuations, whose amplitudes and
frequencies are directly related to properties of the periodic orbits
of the underlying classical dynamics.
Thus, we analyze the fluctuations of the scaled quantum
spectrum~\eqref{eq:scaled_quantum_spectrum} via Fourier transform in
the variable $n_{\mathrm{eff}}$, i.e., a quantum recurrence spectrum
is obtained as
\begin{equation}
  C(S) = \frac{1}{2\pi} \int \varrho(n_{\mathrm{eff}}) e^{-i\tilde S n_{\mathrm{eff}}/\hbar} \dd n_{\mathrm{eff}} \, .
  \label{eq:quantum_recurrence_spectrum}
\end{equation}
The quantum recurrence spectrum~\eqref{eq:quantum_recurrence_spectrum}
should provide peaks at frequencies given by the scaled actions
$\tilde S_{\mathrm{po}}$ of the periodic orbits of the associated
classical exciton dynamics.
Due to the finite number of converged states obtained from numerically 
solving the generalized eigenvalue problem~\eqref{eq:H_scal} the peaks
appear broadened in comparison to the full (infinite) spectrum.
This can also be understood in the following way. The finite spectrum can be obtained 
by multiplying the infinite one with a rectangular window function.
Fourier transforming this expression will give the convolution of the 
delta peaks of the Fourier transformed infinite spectrum and
\begin{equation}
	\frac{\sin(\Delta n_\mathrm{eff} \tilde S_{\mathrm{po}}/(2\hbar))}{\pi \tilde S_{\mathrm{po}}/\hbar} e^{-i n_\mathrm{eff}^0 \tilde S_{\mathrm{po}}/\hbar}\, ,
\end{equation}
with the length of the finite spectrum $\Delta n_{\mathrm{eff}}$ 
and its center $n_\mathrm{eff}^0$.
In addition to the main peaks this will also lead to the appearance of side peaks.
To suppress these unwanted features we use a Gaussian window function
\begin{equation}
  w(n_\mathrm{eff}) \equiv \exp(-\frac{(n_{\mathrm{eff}}-n_{\mathrm{eff}}^0)^2}{2\sigma^2}) \, ,
\end{equation}
where we choose $\sigma \approx \Delta n_\mathrm{eff}/6$.
\begin{figure*}
  \includegraphics[width=0.92\textwidth]{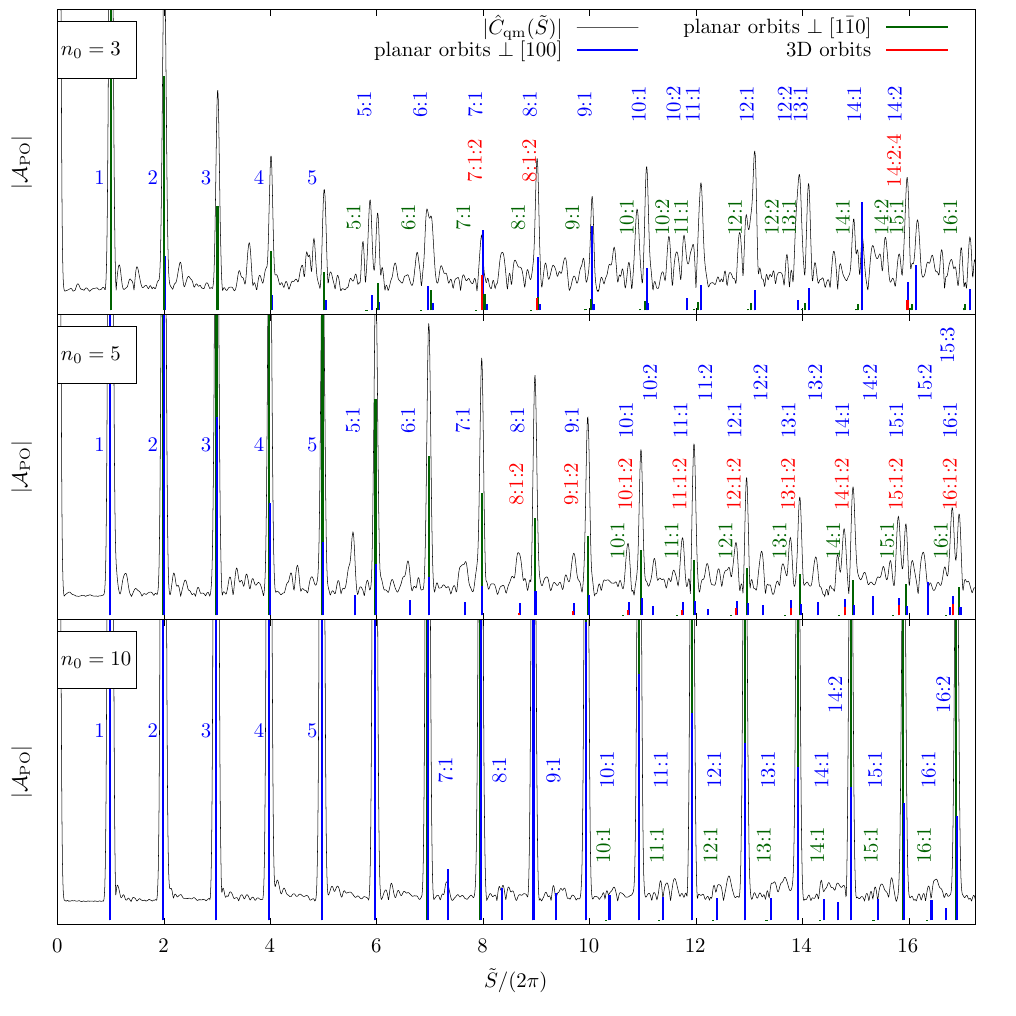}
  \caption{Recurrence spectra for $n_0=3$ (top), $n_0=5$ (middle), and $n_0=10$ (bottom).
    The semiclassical amplitudes are shown as colored bars. Amplitudes 
    for orbits in the plane normal to $[100]$ are shown in blue, 
    amplitudes for orbits in the plane normal to $[1\bar{1}0]$ in green 
    and amplitudes for three-dimensional orbits are shown in red. The amplitudes
    are labeled with the winding numbers of the corresponding orbits.
    The quantum recurrence spectra (black line) are shifted upwards 
    for better visibility. it coupling are valid for the
    corresponding individual plots.
    The recurrence spectrum for $n_0=5$ has already been shown in Ref.~\cite{Ertl2022Signatures}.}
  \label{fig:amplitudes}
\end{figure*}
The resulting expression for the quantum mechanical recurrence spectrum is given by
\begin{align}
	&\hat{C}(\tilde S) = \frac{1}{2\pi} \sum_{k=1}^{k_\mathrm{max}} \int w(n_\mathrm{eff}) \delta(n_\mathrm{eff} - n_{\mathrm{eff},k}) e^{-i n_\mathrm{eff} \tilde S/\hbar}~\mathrm{d}n_\mathrm{eff}\nonumber\\
	&= \frac{1}{2\pi} \sum_{k=1}^{k_\mathrm{max}} w(n_{\mathrm{eff},k})\left[ \cos(n_{\mathrm{eff},k} \tilde S/\hbar) - i\sin(n_{\mathrm{eff},k} \tilde S/\hbar) \right]\, ,
\end{align}
where $k_\mathrm{max}$ is the number of converged eigenvalues $n_{\mathrm{eff},k}$ considered.

\subsubsection{Semiclassical analysis and discussion}
The quantum recurrence spectra obtained from the spectra in
Fig.~\ref{fig:spectra} are shown as solid black lines in
Fig.~\ref{fig:amplitudes} for the three different values of $n_0$.
They exhibit distinct peaks at certain values of the scaled action
$\tilde S/(2\pi)$.
The number of peaks and thus the complexity of the quantum recurrence
spectra increases with decreasing values of $n_0$, i.e., with
decreasing energy of the excitons.
The observed structures in the quantum recurrence spectra can be
explained and interpreted with the help of the periodic exciton orbits.
The semiclassical amplitudes of periodic orbits at positions
$\tilde S_{\mathrm{po}}$ are shown in Fig.~\ref{fig:amplitudes} as
colored bars, labeled by the winding numbers of the  corresponding
orbits.
The majority of the peaks in the quantum recurrence spectra can be understood
in terms of classical orbits. 
For low actions $\tilde S/(2\pi) $ all peaks in the recurrence spectra can
be assigned to the nearly circular orbits, which appear as central fixed point
in the two symmetry planes in Fig.~\ref{fig:psos_big}. 
These orbits move on one-dimensional tori and can therefore be labeled
by an integer winding number $M_1$ which characterizes the repetitions
of the orbits.

 For increasing actions longer two-dimensional and three-dimensional orbits start
 to contribute to the recurrence spectra, leading to an increased density of peaks.
 These orbits belong to motion on two-dimensional tori in the two symmetry planes
 characterized by winding numbers $M_1{:}M_2$ or motion on fully three-dimensional
 tori characterized by winding numbers $M_1{:}M_2{:}M_3$.
 Orbits with the same winding numbers $M_1{:}M_2$ appear clustered together 
 in the recurrence spectra in Fig.~\ref{fig:amplitudes}. Therein orbits in 
 the mostly stable symmetry plane normal to $[100]$ exhibit the highest action 
 as well as semiclassical amplitudes. The orbits in the unstable symmetry plane normal
 to $[1\bar{1}0]$ have the lowest action in the cluster and typically also the lowest
 semiclassical amplitude. At intermediate actions the three-dimensional orbits can be found. 
 
For $n_0=3$ only two pairs of primitive three-dimensional orbits (not
counting repetitions) are found in the presented action range.
Due to their small number the value $g''_E$ for the corresponding
orbits in the plane normal to $[100]$ are used for the calculation of
their semiclassical amplitudes at $n_0=3$.
This should provide a good approximation since the action of these
orbits approaches their two-dimensional counterpart.
Again, their $M_1{:}M_2$ values correspond to the region in which a
change in stability properties can be observed in
Fig.~\ref{fig:Stability_plane} (top).  
For $n_0=5$ multiple three-dimensional orbits exist in the given
range, whereas for $n_0=10$ the three-dimensional orbits start to
appear only at higher actions.
For increasing $n_0$ the contributions of the two- and
three-dimensional orbits decrease at low actions $\tilde S/(2\pi)$
compared to the one-dimensional orbits.
On the one hand the relative amplitudes of orbits with the same
winding numbers decrease for increasing $n_0$.
On the other hand the range of $M_1{:}M_2$ values increases with $n_0$ yielding
more contributions for higher actions $\tilde S/(2\pi)$.
The maximum value $(M_1/M_2)_\mathrm{max}$ is reached at the nearly circular
one-dimensional orbits. Since $(M_1/M_2)_\mathrm{max}$ increases
for increasing $n_0$ the secular motion slows down in the neighborhood of the
central elliptical fixed points giving a more hydrogenlike behavior in this region.
This also becomes apparent in the recurrence spectra. 
In the hydrogenlike case only Kepler ellipses with scaled action
$\tilde S_{\mathrm{po}}/(2\pi) = n$ with $n=1,2,\dots$
exist and therefore the peaks in the corresponding recurrence spectrum are located
at the integers giving the number of repetitions of the orbits.
When introducing the energy surface for the yellow series $W_1$ only
the nearly circular orbits in the symmetry planes are periodic after one cycle
providing a similar contribution like the Kepler ellipses in the hydrogenlike case.
In comparison, contributions of other two-dimensional orbits in the
symmetry planes and three-dimensional orbits become important for
larger values of $\tilde S/(2\pi)$.
With increasing values of $n_0$ this effect becomes more prominent and
leads to a more hydrogenlike appearance of the recurrence spectra
at $n_0=5$ and $10$ in Fig.~\ref{fig:amplitudes}.
This trend can also be observed at higher values of $n_0$.

\section{Conclusion and outlook}
\label{sec:conclusion}
In this paper we investigated the classical dynamics of the yellow
excitons in cuprous oxide and the contributions of classical periodic
exciton orbits to the quantum spectra at various energies.
This was achieved by applying the scaling technique for the spin-orbit 
coupling introduced in Ref.~\cite{Ertl2022Signatures}.
For the two distinct symmetry planes normal to the $[100]$ axis and
the $[1\bar{1}0]$ axis the \acp{PSOS} revealed the phase space
structure, viz.\ the existence of a central fixed point surrounded by
near-integrable tori and a small chaotic region.
The secular motion of orbits around the central fixed points becomes
slower and the size of the chaotic regions decreases with increasing
energy.
In accordance with the Poincar\'e-Birkhoff theorem periodic orbits on
the near-integrable tori exist in pairs and can be labeled by two
integer winding numbers $M_1{:}M_2$.
Out of the symmetry planes fully three-dimensional orbits on resonant
tori characterized by three winding numbers $M_1{:}M_2{:}M_3$ do exist.

The existence of classical periodic exciton orbits has been verified
directly in quantum mechanical exciton spectra.
The obtained results go significantly beyond those presented in
  Ref.~\cite{Ertl2022Signatures}, where the analysis has been restricted to a
  single energy value with $n_0=5$.
  Here, we have extended our analysis to different energies and discussed
  the energy dependence of the classical exciton dynamics and the corresponding
  recurrence spectra by way of examples at $n_0=3$, $5$, and $10$.
The Fourier transform quantum recurrence spectra show detailed
structures of peaks located at distinct values of the scaled action.
Their occurrence can both qualitatively and quantitatively be
explained in terms of periodic exciton orbits by application of trace
formulas from semiclassical periodic orbit theories~\cite{Gut90,Berry76,Ullmo96}.
Line by line comparisons reveal a good agreement between semiclassical
and quantum mechanical recurrence spectra at various energies.
The recurrence spectra show an increasing complexity with decreasing
energy, where two- and three-dimensional periodic orbits occur at
lower values of the scaled action and with increased amplitudes 
compared to the nearly circular orbits. We thus observe significant 
deviations from a purely hydrogenlike system where the Kepler 
ellipses and their repetitions would provide peaks only at multiples 
of $\tilde S= 2\pi$.
This is related to a growing influence of the spin-orbit
interaction in the cuprous oxide semiconductor and thus an increasing
velocity of the secular motion of the exciton orbits compared to
hydrogenlike Keplerian orbits.

In this paper we have focused on exciton spectra and the classical
exciton dynamics related to the yellow exciton series in cuprous oxide.
In future work it would be interesting to further investigate the
classical dynamics of excitons in cuprous oxide, including the green
exciton series~\cite{Rommel2020Green}.
The PSOS are only capable of presenting the phase space of a two-dimensional
system and are therefore not suited to study the phase space of excitons
when including three-dimensional motion. 
One approach to study the corresponding dynamics could be using Lagrangian 
descriptors~\cite{PhysRevLett.105.038501,MANCHO20133530,DAQUIN2022133520}, 
which have proven to be a useful tool for revealing phase structures in 
non-integrable systems.
 
Another interesting topic would be to extend the semiclassical treatment.
In this work the contributions of classical orbits were studied by 
connecting classical orbits to the peaks in the quantum recurrence 
spectra. Due to the large regular part of phase space and the possibility
of reconstructing action variables from the classical orbits,
 it will be interesting to see if the spectrum of excitons in cuprous oxide
can be directly connected to classical orbits using the
EBK-quantization method~\cite{Curtis2004}.
These approaches have successfully been applied for the hydrogen atom
in external electric and magnetic fields~\cite{Gekle2006a,Gekle2007a}.
This would allow for a direct understanding of the quantum spectra in
terms of classical exciton orbits.\\

\acknowledgments
This work was supported by Deutsche Forschungsgemeinschaft (DFG)
through Grant No.~MA1639/16-1.

%

\end{document}